\documentclass[%
 reprint,]{revtex4-1}
\usepackage[colorlinks,
            linkcolor=blue,
            anchorcolor=blue,
            citecolor=blue
            ]{hyperref}
\usepackage{graphicx}
\usepackage{dcolumn}
\usepackage{bm}
\usepackage{mathrsfs}
\usepackage{amssymb}

\begin{document}

\preprint{APS/123-QED}

\title{Discussion on Lorentz Invariance Violation of Non-commutative Field Theory and Neutrino Oscillation}
\author{Cui-Bai Luo$^{1}$, Song Shi$^{1}$, Yi-Lun Du$^{1}$, Yong-Long Wang$^{1}$}
\author{Hong-Shi Zong$^{1,2,3}$} \email{zonghs@nju.edu.cn}
\address{$^{1}$Department of Physics, Nanjing University, Nanjing 210093, China}
\address{$^{2}$Joint Center for Particle, Nuclear Physics and Cosmology, Nanjing 210093, China}
\address{$^{3}$State Key Laboratory of Theoretical Physics, Institute of Theoretical Physics, CAS, Beijing 100190, China}

\date{\today}

\begin{abstract}
Depending on deformed canonical anti-commutation relations, massless neutrino oscillation based on Lorentz invariance violation in non-commutative field theory is discussed. It is found that the previous studies about massless neutrino oscillation within deformed canonical anti-commutation relations should satisfy the condition of new Moyal product and new non standard commutation relations. Furthermore, comparing the Lorentz Invariant Violation parameters $\mathit{A}$ in the previous studies with new Moyal product and new non standard commutation relations, we find that the orders of magnitude of non-commutative parameters (Lorentz invariant Violation parameters $\mathit{A}$) is not self-consistent. This inconsistency means that the previous studies of Lorentz invariance violation in non-commutative field theory may not naturally explain massless neutrino oscillation. In other words, it should be impossible to explain neutrino oscillation by Lorentz invariance violation in non-commutative field theory. This conclusion is supported by the latest atmospheric neutrinos experimental results from the Super-Kamiokande Collaboration, which show that no evidence of Lorentz invariance violation on atmospheric neutrinos was observed.

\begin{description}
\item[PACS numbers]
11.30.Cp, 14.60.St, 11.10.Nx
\end{description}
\end{abstract}

\maketitle

\section{\label{sec:level1}Introduction}
Neutrino oscillation was proposed by Pontecorvo \cite{Pontecorvo,PRD172369} to explain the deficit of solar and atmospheric neutrinos \cite{PRD64093005,PRL844035} in fluxes measured on earth. It is well known that this phenomenon is beyond the standard model of particle physics. In order to explain neutrino oscillation, the conventional scenario is to assume that the neutrino mass is not null, and the main theory is the Seesaw Mechanism. Such an assumption is extremely important to particle physics, astrophysics and cosmology. Among the six essential parameters for describing neutrino oscillation, the $CP$ phase $\delta$ is still unknown.
The Daya Bay experiment measured the mixing angle $\textrm{sin}^22\theta_{13} = 0.092 \pm 0.017$, which may provide a guideline for the future development of neutrino physics, in particular for the understanding of the absence of anti-matter in the universe. The latest experiments from IceCube Collaboration show that no evidence for sterile neutrinos was observed \cite{PhysRevLett117071801}, which poses a new challenge to how to establish a new theory of neutrino oscillations.

Another theory that explains neutrino oscillations is the non-commutative field and Lorentz Invariance Violation theory. The issue of massless neutrino oscillation in non-commutative field and Lorentz Invariance Violation has been extensively discussed in literatures \cite{PhysRevD69016005, arias2006cptlorentz, PhysRevD80123522, PhysRevD85016013, ma2009, PhysRevD91036009, Arias2006kk, Carmona2003222, Coleman1997249}. S. Coleman and S.L. Glashow \cite{Coleman1997249} noted that neutrino oscillations may take place for massless neutrinos in which case the Lorentz invariance is violated in the neutrino sector. They argued that the existence of high-energy cosmic rays places strong constraints on Lorentz non-invariance. Furthermore, if the maximum attainable speed of a particle depends on its identity, then neutrinos, even if massless, may exhibit flavor oscillations \cite{Coleman1997249}.

In astronomical phenomena, possible violations of Lorentz invariance have been investigated for a long time using the observed spectral lags of gamma-ray bursts (GRBs) \cite{Ellis201350, PhysRevLett110201601, Wei2016A, Zhang2015108}. In string theories, Lorentz symmetry breakdown is natural when the perturbative string vacuum is unstable \cite{RevModPhys73977}. Another approach of introducing the Lorentz symmetry breakdown is the non-commutative field theory, which is found to be physically equivalent to a subset of a general Lorentz-violating Standard-Model extension involving ordinary fields \cite{Guralnik2001450,PhysRevLett87141601}. Moreover, this kind of symmetry breakdown may also arise from quantum gravity \cite{PhysRevD66124006,PhysRevLett89231301} or field theories with gravity \cite{PhysRevD39683}. Topological defect in the space-time, like the cosmic string \cite{Kibble}, is considered to be through phase transitions during the evolution of the Universe, which involves Lorentz symmetry breaking.

In this work, we discuss previous studies about massless neutrino oscillation in non-commutative field and CPT/Lorentz invariance violation. Section \ref{sec:level2} shows that in conventional non-commutative fields, for neutrino, there are not non-commutative effects and massless neutrino oscillation. In section \ref{sec:level3}, we introduce the results of previous studies, namely that massless neutrino oscillation conditions must be consistent with deformed canonical anti-commutation relations Eq. (\ref{ncf1}). By means of the conventional noncommutative field one can not derive the Eqs. (\ref{la}, \ref{ncf1}), and we stress that in the non-commutative field, the previous studies did not consider the correct condition in which the deformed canonical anti-commutation relations (\ref{ncf1}) are established. In section \ref{sec:level4}, we find that the deformed canonical anti-commutation relations (\ref{ncf1}) should satisfy the new Moyal product (\ref{star2}) and new non standard commutation relations (\ref{nnscr}). Finally,  it is found that there is still uncertainty in the previous studies about massless neutrinos. If massless neutrino oscillations could take place in the case of Lorentz invariance violation in new non-commutative field (new Moyal product), the orders of magnitude of non-commutative parameters and Lorentz Invariance Violation parameters are still inconsistent with previous data.

\section{\label{sec:level2}Conventional non-commutative fields}
The idea that the space-time coordinates do not commute is quite old in mathematics and physics \cite{PhysRev7138,  alainconnes}. Since the discovery in string theory that the low-energy effective theory of a D-brane in the background of a Neveu-Schwarz-Neveu-Schwarz B field lives on a non-commutative space \cite{seibergwitten}, the issue of non-commutative geometry has undergone a recent revival and has been extensively discussed \cite{szabo, ihab, PhysRevLett94151602, PhysRevLett87141601, mmsheikh}. It is well known that we can construct a non-commutative (NC) quantum field theory by replacing ordinary fields with non-commutative fields and ordinary products with Groenewold-Moyal star product \cite{moyal,szabo,Bayen197861} defined by
\begin{eqnarray}\label{star}
f(x)\star g(x) &=& f(x)exp(\frac{i}{2}\stackrel{\leftarrow}{\partial_i}\theta^{ij}\vec{\partial}_j)g(x) \nonumber \\
&=& f(x)g(x)+\sum^{\infty}_{n=1}(\frac{i}{2})^n\frac{1}{n!}\theta^{i_1j_1}\cdots \theta^{i_nj_n} \nonumber \\
&& \cdot \partial_{i_1}\cdots\partial_{i_n}f(x)\partial_{j_1}\cdots \partial_{j_n}g(x).
\end{eqnarray}
Considering the gauge theories such as quantum electrodynamics (QED), the Hermitian Lagrangian is
\begin{eqnarray} \label{nqed}
\mathcal{L}_N=\frac{i}{2}\bar{\hat{\psi}}\gamma^\mu\star \stackrel{\leftrightarrow}{\hat{D}}_\mu \hat{\psi} - m\bar{\hat{\psi}} \star \hat{\psi} -\frac{1}{4}F_{\mu\nu}\star F^{\mu\nu},
\end{eqnarray}
where $\hat{\psi}$ is the fields in non-commutative QED space, and $\psi$ is the corresponding fields for conventional QED space.
Submitting the Seiberg-Witten map
\begin{eqnarray} \label{swm}
\hat{A}_\mu &=& A_\mu -\frac{1}{2}\theta^{\alpha \beta}A_\alpha(\partial_\beta A_\mu + F_{\beta \mu}), \nonumber \\
\hat{\psi}&=&\psi - \frac{1}{2}\theta^{\alpha \beta}A_\alpha \partial_\beta \psi
\end{eqnarray}
into Eq. (\ref{nqed}) and applying the definition (\ref{star}), one can obtain the ordinary quantum field theory
that is physically equivalent to non-commutative QED to leading order in $\theta_{\mu \nu}$
\begin{eqnarray}\label{new}
\mathcal{L}_N&=&\frac{1}{2}i\epsilon_1\bar{\psi}\gamma^\mu \stackrel{\leftrightarrow}{D}_\mu \psi - m\epsilon_2\bar{ \psi}\psi  -\frac{1}{4}\epsilon_3 F_{\mu\nu}F^{\mu\nu} + \epsilon_4,
\end{eqnarray}
where
\begin{eqnarray}
\epsilon_1 &=&\epsilon_2=1-\frac{1}{4}q\theta^{\alpha \beta}F_{\alpha \beta}, \quad \epsilon_3=2\epsilon_1 -1, \nonumber \\
\epsilon_4 &=& \frac{1}{2}q\theta^{\alpha \beta}F_{\alpha \mu}[\frac{1}{2}i\bar{\psi}\gamma^\mu \stackrel{\leftrightarrow}{D}_\beta \psi - F_{\beta\nu}F^{\mu\nu}].
\end{eqnarray}
The Lagrangian (\ref{new}) contains CPT violation term, which consists of ordinary QED plus non-renormalizable Lorentz violating corrections. The Lagrangian is also manifestly gauge invariant. Here it should be noted that all non-commutative effects vanish for neutral fermions ($q=0$). Due to the properties of the Moyal product (\ref{star}), one concludes that free field theory cannot be modified by space-time NC \cite{szabo}. In addition, Chaichian and Pre\v snajder \textit{et al}. \cite{Chaichian2002132} presented the no-go theorem: this theorem shows that matter fields in the non-commutative $U(1)$ gauge theory can only have $\pm 1$ or $0$ charges, and for a generic non-commutative $\Pi_{i=1}^n(N_i)$ gauge theory, matter fields can be charged under at most two of the $U(N_i)$ gauge group factors. To sum up, for neutral fermions in conventional non-commutative fields, there are no non-commutative effects and massless neutrino oscillation.

\section{\label{sec:level3}Deformed anti-commutation relations and neutrino oscillation}
Considering the above situation, massless neutrino oscillation based on CPT/Lorentz invariance violation were brought forward by means of deformed equal-time anti-commutation relations or Lorentz-violating extension of the standard model \cite{PhysRevD69016005,arias2006cptlorentz,PhysRevD80123522,PhysRevD85016013,ma2009,PhysRevD91036009,Arias2006kk,Carmona2003222}. In Lorentz-violating extension of the standard model, the general equations of motion for free propagation can be written as a first-order differential operator acting on the object $\nu_\beta$
\begin{eqnarray}\label{lvl}
&&(i\Gamma^\nu_{\alpha \beta}\partial_\nu - M_{\alpha \beta})\nu_\beta=0, \nonumber \\
M_{\alpha \beta}& :=&m_{\alpha \beta} + im_{5\alpha \beta}\gamma_5 + a^{\mu}_{\alpha \beta}\gamma_\mu + b^{\mu}_{\alpha \beta}\gamma_5\gamma_\mu +\frac{1}{2}H^{\mu \nu}_{\alpha \beta}\sigma_{\mu \nu},\nonumber \\
\Gamma^\nu_{\alpha \beta}&:=& \gamma^\nu\delta_{\alpha \beta} +c^{\mu \nu}_{\alpha \beta}\gamma_\mu + e^{\nu}_{\alpha \beta} +if^{\nu}_{\alpha \beta}\gamma_5 + \frac{1}{2}g^{\lambda \mu \nu}_{\alpha \beta}\sigma_{\lambda \mu},
\end{eqnarray}
where $m_5$ is Lorentz and \textit{CPT} conserving, the coefficients $c, d, H$ are \textit{CPT} conserving but Lorentz violating, while $a, b, e, f, g$ are both \textit{CPT} and Lorentz violating.
In Refs. \cite{PhysRevD69016005,PhysRevD85016013} the mass $m_{\alpha \beta}$ of neutrinos are considered, in this section we only discuss massless neutrino, so we won't discuss this issue here.

For simplicity, from Eq. (\ref{lvl}) we restrict ourself to the part of the Lagrangian given by Ref.  \cite{arias2006cptlorentz}
\begin{eqnarray}\label{la}
\mathcal{L}=i\bar{\nu}_\alpha \gamma^\mu \partial_\mu \nu_\alpha +i\bar{\nu}_\alpha c^{\mu \nu}_{\alpha \beta}\gamma_\mu \partial_\nu  \nu_\beta - a^{\mu}_{\alpha \beta}\bar{\nu}_\alpha \gamma_\mu \nu_\beta,
\end{eqnarray}
with the constant NC parameter or Lorentz violating parameter $c^{\mu \nu}_{\alpha \beta}$, where the subscript $\alpha= {e, \mu, \tau}$ runs over the flavor quantum number, and $\nu_\alpha$ is the fields in non-commutative space. In this work we had assumed $a^\mu_{\alpha \beta}= 0$, which means that one only consider the case of \textit{CPT} conserving but Lorentz violating ($c^{\mu \nu}_{\alpha \beta}$).

The dimensionless coefficients $c^{\mu \nu}_{\alpha \beta}$ can have both symmetric and antisymmetric parts but can be assumed traceless. Then by Lagrangian (\ref{la}) we obtain
\begin{eqnarray}\label{ncm}
\Pi_{\nu_\alpha}=iC_{\alpha \beta}\nu^{\dag}_\beta=i(\delta_{\alpha \beta}+c^{00}_{\alpha \beta})\nu^{\dag}_\beta.
\end{eqnarray}
This leads to the deformed canonical anti-commutation relations
\begin{eqnarray}\label{ncf1}
\{\nu_\alpha(x), \nu^\dag_\beta(y)\}= A_{\alpha \beta}\delta(x-y),
\end{eqnarray}
where
\begin{eqnarray} \label{aandc}
A_{\alpha \beta}=(\delta_{\alpha \beta} + c_{\alpha \beta}^{00})^{-1}.
\end{eqnarray}

With the basic vector $\nu_L(p), \nu_R(-p)$, by the Lagrangian (\ref{la}) we have the Hamiltonian matrix for left-handed neutrinos and right-handed anti-neutrinos
\begin{eqnarray}\label{}
\mathcal{H}_{\alpha \beta}=| \vec{p}|\delta_{\alpha \beta} +c^{\mu \nu}_{\alpha \beta}\frac{p_\mu p_\nu}{| \vec{p}|} \pm a^\mu_{\alpha \beta} \frac{p_\mu }{| \vec{p}|}.
\end{eqnarray}
For unitary matrix $U$, which diagonalizes $\mathcal{H}$, the diagonalized eigenenergy matrix and neutrino energy eigenstates are respectively
\begin{eqnarray}
E_{ij}=U^\dag_{i\alpha}\mathcal{H}_{\alpha \beta}U_{\beta j}, \quad |\nu_i  \rangle= U^\dag_{i\alpha}|\nu_{\alpha} \rangle,
\end{eqnarray}
where $\nu_{i}$ and $\nu_{\alpha}$ represent different energy eigenstates and the flavor eigenstates. Due to matrix $U$ being unitary, the energy eigenstates can be transformed to
\begin{eqnarray}
|\nu_\alpha  \rangle=\Sigma_{i\beta}(U^\dag_{i\alpha})^*e^{-iE_it}U^\dag_{i\beta}|\nu_{\beta} \rangle.
\end{eqnarray}
When one only consider the case of \textit{CPT} conserving but Lorentz violating, one can get the oscillation probability $P_{\nu_\alpha \to \nu_\beta}$
\begin{eqnarray} \label{zwzzd}
P_{\nu_e \to \nu_\mu} &=&\frac{d^2}{d^2+b^2+c^2}\mathrm{sin}[(\frac{\sqrt{d^2+b^2+c^2}}{2})L] \nonumber \\
P_{\nu_\mu \to \nu_\tau} &=&\frac{b^2}{d^2+b^2+c^2}\mathrm{sin}[(\frac{\sqrt{d^2+b^2+c^2}}{2})L] \nonumber \\
P_{\nu_\mu \to \nu_\mu} &=&1- \frac{d^2 +b^2}{a^2+b^2+c^2}\mathrm{sin}[(\frac{\sqrt{d^2+b^2+c^2}}{2})L] ,
\end{eqnarray}
where $d=2 c^{00}_{e\mu}E, b=2 c^{00}_{\mu \tau}E$ and $ c =c^{00}_{\mu \mu}E$.

Based on the above analysis and the experimental data from K2K, KamLAN, MiniBoone and LSND experiments, the data of oscillation probability $P_{\nu_\alpha \to \nu_\beta}$, $E, L$, \textit{et al} can be given. Then the elaborate bounds on the non-commutative parameters or Lorentz Violation parameters $c^{00}_{\alpha \beta}$ and $A_{\alpha \beta}$ within the non-commutative filed model are obtained. The values of Lorentz Violation parameters $A_{\alpha \beta}$ are shown in Table~\ref{table1} \cite{arias2006cptlorentz,ma2009,Arias2006kk,RevModPhys8311}.

\begin{table}[tbp]
\centering
\begin{tabular}{|lr|c|}
\hline
Parameters & Classes & Estimated Value  \\
\hline
$A_{e\mu}$ & Solar neutrino & $\sim10^{-17}, 10^{-18}$ \\
$A_{\mu \tau}$ & Atmospheric neutrino & $\sim10^{-22}, 10^{-23}$ \\
$A_{\alpha \beta}$ & ---& $\sim 10^{-20}$ \\
\hline
\end{tabular}
\caption{\label{table1}Estimated Value \cite{arias2006cptlorentz,ma2009,Arias2006kk,RevModPhys8311}}
\end{table}

In the above analysis, we show that massless neutrino oscillation does not occur in the conventional non-commutative field, unless this is in the deformation of the canonical anti-commutation relations (\ref{ncf1}). If the canonical anti-commutation relations (\ref{ncf1}) holds, we have the oscillation probability $P_{\nu_\alpha \to \nu_\beta}$ such as Eq. (\ref{zwzzd}),  then the Lorentz Violation parameters $c^{00}_{\alpha \beta}$ and $A_{\alpha \beta}$ are obtained. However, the previous studies did not consider the correct condition in which the relations (\ref{ncf1}) were established. If the conventional anti-commutation relationship is deformed, then we may have to consider the deformation relations between $\nu_\alpha$ and $\nu_\beta$ or $\Pi_\alpha$ and $\Pi_\beta$. So, in the following section, by a new Moyal product we will derive the proper condition on which the relations (\ref{ncf1}) can be established.

\section{\label{sec:level4}New Moyal product and new non standard commutation relations}
Refs. \cite{Arias2006kk, Carmona2003222}, introduce a new kind of noncommutativity in field theory. The simplest option is to deform the commutator of fields in analogy with the deformation of the commutator of the coordinates. In order to preserve the locality in the new set of canonical commutation relations, for a complex scalar field, non standard commutation relations become
\begin{eqnarray}\label{nccr}
\quad [\phi_i(t,\vec{x}), \phi_j(t,\vec{y})]&=&i\bar{\theta}_{ij} \delta(\vec{x},\vec{y}), \nonumber \\
\quad [\phi_i(t,\vec{x}),\Pi_j(t,\vec{y})]&=& i\delta_{ij} \delta(\vec{x},\vec{y}),
\end{eqnarray}
where the NC parameter $\bar{\theta}$ has the dimension of $\sqrt{\theta}$, and $\theta$ is the usual NC parameters in NC space

The Moyal product Eq. (\ref{star}) allows mapping of the study of non-commutative field theories into that of ordinary field theories, where the ordinary product is replaced by the star product \cite{alainconnes}. J.M. Carmona \textit{et al} \cite{Carmona2003222} propose a new Moyal product which is completely different from that of Eq. (\ref{star}), such as
\begin{eqnarray}\label{star2}
\Phi_1(\phi)\star \Phi_2(\phi) = \lim_{\eta,\zeta \to \phi}e^{\frac{i}{2}\bar{\theta}\epsilon_{ij}\int dx \frac{\delta}{\eta_i(x)} \frac{\delta}{\zeta_j(x)}}\Phi_1(\eta) \Phi_2(\zeta). \nonumber \\
\end{eqnarray}
The new Moyal product between functionals is consistent with the commutation relations (\ref{nccr}), and the standard properties of the Moyal product also hold as
\begin{eqnarray}\label{starnc}
 \phi_i(x) \star \phi_j(x') &=&\phi_i(x) \phi_j (x')+\frac{i}{2}\bar{\theta}\epsilon_{ij}\delta(x-x') \nonumber \\
\quad [\phi_i(x),\phi_j(x')]_{\star}  &=& \phi_i(x) \phi_j (x') - \phi_j(x') \phi_i (x)  \nonumber \\
&&+ \frac{i}{2}\bar{\theta}\epsilon_{ij}\delta(x -x') - \frac{i}{2}\bar{\theta}\epsilon_{ji}\delta(x -x')\nonumber \\
&=&i\bar{\theta}_{ij} \delta(\vec{x},\vec{y}).
\end{eqnarray}
In this work, we always discuss the problem in the non-commutative space, so for convenience we ignore the product notation "$\star$", then the above expression becomes Eq. (\ref{nccr}).

Similarly, under this definition (\ref{star2}), we try to generalize non-standard commutation relations of complex scalar field (\ref{nccr})  to fermion field,
\begin{eqnarray}\label{nccr2}
\{\psi_i(t,\vec{x}), \psi_j(t,\vec{y})\}&=&i\bar{\theta}_{ij} \delta(\vec{x},\vec{y}), \nonumber \\
\{\psi_i(t,\vec{x}),\Pi_j(t,\vec{y})\}&=& i\delta_{ij} \delta(\vec{x},\vec{y}),
\end{eqnarray}
and
\begin{eqnarray}\label{starnc2}
\{\psi_i(t,\vec{x}),\psi_j(t,\vec{y})\}_{\star}  =i\bar{\theta}_{ij} \delta(\vec{x},\vec{y}).
\end{eqnarray}

Now let's discuss how to get neutrino' deformed canonical anti-commutation relations (\ref{ncf1}), under the new Moyal product (\ref{star2}) and new set of canonical commutation relations (\ref{nccr2}, \ref{starnc2}). The fact that the neutrinos conjugate momentum (\ref{ncm}) can be  demonstrated with the help of the canonical anti-commutation relations (\ref{ncf1}), which may also be used to define a new canonical anti-commutation relations of conjugate momentum. One can do the same trick here as with the derivation of Eq. (\ref{ncm}). Let us define
\begin{eqnarray}\label{wz}
\Pi_\alpha =iC_{\alpha \beta}\nu^{\dag}_\beta, \quad (C_{\alpha \beta})^{-1}= A_{\alpha \beta}.
\end{eqnarray}
Inserting the equation above into  non standard commutation relations (\ref{nccr2}, \ref{starnc2}), then we have
\begin{eqnarray}\label{nscr}
\{\nu_\alpha(t,\vec{x}),\nu^{\dag}_\beta(t, \vec{y})\}&=&A_{\alpha \beta} \delta(\vec{x},\vec{y}), \nonumber \\
\{\Pi_\alpha (t,\vec{x}),\Pi_\beta (t,\vec{y})\}&=&iC^2\bar{\theta}^{\dag}_{\alpha \beta} \delta(\vec{x},\vec{y})
\end{eqnarray}
Under the conditions of new Moyal product, obviously, Eq. ({\ref{nscr}) implies that we have to generalize non-standard commutation relations (\ref{nccr2}) to the new non-standard commutation relations
\begin{eqnarray}\label{nnscr}
\{\nu_\alpha (t,\vec{x}),\nu^{\dag}_\beta(t,\vec{y})\}&=&A_{\alpha \beta} \delta(\vec{x},\vec{y}), \nonumber \\
 \{\nu_\alpha (t,\vec{x}), \nu_\beta(t,\vec{y})\}&=&i\bar{\theta}_{\alpha \beta} \delta(\vec{x},\vec{y}), \nonumber \\
\{\nu_\alpha(t,\vec{x}),\Pi_\beta(t,\vec{y})\}&=& i\delta_{\alpha \beta} \delta(\vec{x},\vec{y}),\nonumber \\
\{\Pi_\alpha (t,\vec{x}), \Pi_\beta(t,\vec{y})\}&=&i\bar{\eta}_{\alpha \beta} \delta(\vec{x},\vec{y}),
\end{eqnarray}
where
\begin{eqnarray} \label{c2}
\bar{\eta} =C^2 \bar{\theta}^{\dag},
\end{eqnarray}
NC parameter $\bar{\eta}$ has the dimension of $\sqrt{\eta}$ such as $\bar{\theta}$ has the dimension of $\sqrt{\theta}$, and $\eta$ is the NC parameter in NC phase space.

Ordinarily, in string theory non-commutativity appears only at the coordinates level $[q_i, q_j] =i\theta_{ij}$ (The violation of the Lorentz symmetry arises directly from the commutator of the coordinates). However, momentum non-commutativity may naturally arise as a consequence of coordinates non-commutativity, as momenta are defined as the partial derivatives of the action with respect to the non-commutative spatial coordinates \cite{singhtp2005,PhysRevD72025010,aef0253}. When non-commutativity of both configuration and momentum spaces is considered, there are the following commutation relations \cite{PhysRevLett87141601,PhysRevD72025010}
\begin{eqnarray}\label{nqnn}
[\hat{x}^{\mu}, \hat{x}^{\nu}] &=&i \theta^{\mu \nu}, \quad  [\hat{p}^{\mu}, \hat{p}^{\nu}] = i\eta^{\mu \nu},  \nonumber \\
\lbrack \hat{x}^{\mu}, \hat{p}^{\nu}]&=&i\hbar  (\delta^{\mu \nu}+\frac{\theta^{\mu \alpha} \eta^{\nu}_{\phantom{i} \alpha}}{4\hbar^2}) \sim i\hbar  \delta^{\mu \nu}.
\end{eqnarray}

Through the above analysis, the deformed canonical anti-commutation relations (\ref{ncf1}, \ref{nscr}) explain massless neutrino oscillation. However, the deformed canonical anti-commutation relations (\ref{ncf1}) should meet new Moyal product (\ref{star2}). Under such conditions, non-standard commutation relations should be generalized to Eq. (\ref{nnscr}), where the NC parameter $(\bar{\theta},\bar{\eta})$ has the dimension of $(\sqrt{\theta}, \sqrt{\eta})$, and $(\theta,\eta)$ are the usual non-commutative parameters in NC space and NC phase space. As a conservative limit, existing experiments bound the scale of the non-commutativity $(\sqrt{\theta}, \sqrt{\eta})$ parameter to  \cite{PhysRevLett87141601,PhysRevD72025010}
\begin{eqnarray}\label{cte}
|\theta| \lesssim (10\mathrm{TeV})^{-2}, \sqrt{\eta} \lesssim 1\mathrm{MeV}/c.
\end{eqnarray}
Submit the value of Eq. (\ref{cte}) into Eqs. (\ref{wz}, \ref{nscr}, \ref{nnscr}) [That means these value $\theta (\theta_{\alpha \beta}), \eta(\eta_{\alpha \beta})$ should satisfy the Eqs. (\ref{wz}, \ref{nscr}, \ref{nnscr})],
so the conditions in Eq. (\ref{aandc}, \ref{c2})
\begin{eqnarray}\label{cte2}
\bar{\eta} \sim \sqrt{\eta} = C^2 \bar{\theta}^{\dag} =(\delta_{\alpha \beta} + c_{\alpha \beta}^{00})^2  \bar{\theta}^{\dag}
 \end{eqnarray}
also hold. Then we can get the solution of $c_{\alpha \beta}^{00}$ by the conditions (\ref{cte}, \ref{cte2}).
So the value of $A_{\alpha \beta}$ can be estimated by formula Eq. (\ref{aandc}) $A_{\alpha \beta} =(C_{\alpha \beta})^{-1}=(\delta_{\alpha \beta} + c_{\alpha \beta}^{00})^{-1}$.
This means that the non-commutative parameter $A$ becomes
\begin{eqnarray}\label{nav}
A (A_{\alpha \beta}) \sim 10^{-10}.
\end{eqnarray}

Comparing with the estimated value of non-commutative parameters as showed in Table~\ref{table1}, obviously, the orders of magnitude of non-commutative parameters or Lorentz Invariance Violation parameters $A(A_{\alpha \beta}) $ in Eq. (\ref{nav}) is inconsistent with previous data (Table~\ref{table1}). In addition, the estimated value $A_{e\mu}$ and $A_{\mu \tau}$ as shown in Table~\ref{table1},  can differ by five orders of magnitude, which is also inconsistent with the results (\ref{nav}).

It is clear now that within this scenario, all the theoretical results are not self-consistent: If the canonical anti-commutation relations (\ref{ncf1}) and the Lagrangian (\ref{la}) hold, then the Lorentz Violation parameters $c^{00}_{\alpha \beta}$ and $A_{\alpha \beta}$ are obtained such as Table~\ref{table1}. Moreover, Eq. (\ref{la}, \ref{ncf1}) should meet the new Moyal product (\ref{star2}) and the new non standard commutation relations (\ref{nnscr}), then the Lorentz Violation parameters $A_{\alpha \beta}$ should meet Eq. (\ref{nav}). However,  the Lorentz Violation parameters $A_{\alpha \beta}$ in Eq. (\ref{nav}) are not consistent with Table~\ref{table1}. This inconsistency means that the previous studies about Lorentz invariance violation in non-commutative field theory may not naturally explain massless neutrino oscillation. In other words, it should be impossible to explain neutrino oscillation by Lorentz invariance violation in non-commutative field theory.

\section{Conclusion}
There are mainly two kinds of  neutrino oscillation theories, one is the Seesaw mechanism, the other is the Lorentz Invariance Violation theory. However, the latest experiments from IceCube Collaboration show that no evidence for sterile neutrinos was observed \cite{PhysRevLett117071801}, which means that the Seesaw theory may face challenges. In this work, we focus on neutrino oscillation in the Lorentz Invariance Violation theory. We find the non-self-consistency of the previous neutrino oscillation studies by Lorentz invariance violation in non-commutative field theory, which means it may not naturally explain massless neutrino oscillation by Lorentz invariance violation in non-commutative field theory.

In this work, depending on deformed canonical anti-commutation relations (\ref{ncf1}), the previous studies about massless neutrino oscillation based on CPT/Lorentz invariance violation were brought forward. However, in conventional non-commutative field theories, since free field theory cannot be modified by space-time NC and all non-commutative effects vanish for neutral fermions, there are not non-commutative effects and massless neutrino oscillation. So just by conventional non-commutative field it can not derive the Eq. (\ref{la}) or deformed canonical anti-commutation relations Eq. (\ref{ncf1}).

The previous study about massless neutrino theory is built on the basis of relations in Eqs. (\ref{la}, \ref{ncf1}), by these relations and neutrino experimental data one gets the Lorentz Violation parameters $A_{\alpha \beta}$ such as in Table~\ref{table1}. However, the previous studies did not consider the correct condition of its establishment (\ref{ncf1}). In this work, we find that the deformed canonical anti-commutation relations (\ref{ncf1}) should satisfy this condition of new Moyal product (\ref{star2}) and non standard commutation relations (\ref{nnscr}), then the Lorentz Violation parameters $A_{\alpha \beta}$ in Eq. (\ref{nav}) hold. However, comparing the Lorentz Violation parameters $A_{\alpha \beta}$ in Eq. (\ref{nav}) with Table~\ref{table1}, the orders of magnitude of non-commutative parameters  $A$ (Lorentz Invariance Violation parameters) are not self-consistent.

Non-self-consistency of results mean that the previous studies about Lorentz invariance violation in non-commutative field theory may not naturally explain massless neutrino oscillation. Therefore, we conclude that it should be impossible to explain neutrino oscillation by Lorentz invariance violation in non-commutative field theory. The latest experimental results from Super-Kamiokande Collaboration about the test of Lorentz invariance with atmospheric neutrinos \cite{PhysRevD91052003}, show that no evidence of Lorentz invariance violation was observed. Our conclusion is support by the latest atmospheric neutrinos experiment. In addition, we want to stress that, possible violations of Lorentz invariance have been investigated for a long time using the observed spectral lags of gamma-ray bursts (GRBs) \cite{Ellis201350, PhysRevLett110201601, Wei2016A, Zhang2015108}, while our conclusion is just the opposite (Because the coefficients of Lorentz invariance violation are not consistent with the experimental data of neutrino oscillations). This implies that the neutrino oscillation theory and possible violations of Lorentz invariance in gamma-ray bursts deserves further study.

\acknowledgments
This work is supported by the National Natural Science Foundation of China (under Grants No. 11274166, No. 11275097, No. 11475085, and No. 11535005), the Jiangsu Planned Projects for Postdoctoral Research Funds (under Grant No. 1401116C), and China Postdoctoral Science Foundation (under Grant No. 2014M561621).

\bibliography{nm}

\end{document}